\documentclass[aps,prd,twocolumn,groupedaddress,showpacs,
a4paper,floats]{revtex4}
\usepackage{graphicx}
\usepackage{amssymb,latexsym}
\usepackage{longtable}

\def\ga{\mathrel{\raise.3ex\hbox{$>$\kern-.75em\lower1ex\hbox{$\sim$}}}}
\def\la{\mathrel{\raise.3ex\hbox{$<$\kern-.75em\lower1ex\hbox{$\sim$}}}}
\def\be{\begin{equation}}
\def\ee{\end{equation}}
\def\ba{\begin{eqnarray}}
\def\ea{\end{eqnarray}}
\def\ga{\mathrel{\raise.3ex\hbox{$>$\kern-.75em\lower1ex\hbox{$\sim$}}}}
\def\la{\mathrel{\raise.3ex\hbox{$<$\kern-.75em\lower1ex\hbox{$\sim$}}}}

\newcommand{\bi}[1]{\bibitem{#1}}
\newcommand{\fr}[2]{\frac{#1}{#2}}

\begin{document}

\preprint{UVIC--TH--2003-05}
\preprint{SUSX-TH/}

\title{Models of quintessence coupled to the electromagnetic field 
and the cosmological evolution of alpha}

\author{E. J. Copeland$^1$, N. J. Nunes$^2$ and M. Pospelov$^{1,3}$}
\affiliation{1. Centre for Theoretical Physics, CPES,
University of Sussex, Brighton BN1~9QJ, UK}
\affiliation{2. School of Mathematical Sciences, Queen Mary, University of
  London, Mile End Road, E1 4NS, UK}
\affiliation{3. Department of Physics and Astronomy,
University of Victoria, 
     Victoria, BC, V8P 1A1 Canada}

\date{\today}

\begin{abstract}
We study the change of the effective fine structure constant
in the cosmological models of a scalar field with a non-vanishing 
coupling to the electromagnetic field. Combining cosmological data 
and terrestrial observations we place empirical constraints on the 
size of the possible coupling and explore a large class of models
that exhibit tracking behavior. The change of the fine 
structure constant implied by the 
quasar absorption spectra together with the requirement of tracking behavior
impose a lower bound of the size of this coupling. Furthermore, the transition 
to the quintessence regime implies a narrow window for this coupling around
$10^{-5}$ in units of the inverse Planck mass. 
We also propose a non-minimal coupling between electromagnetism and 
quintessence which has the effect of leading only 
to changes of alpha determined 
from atomic physics phenomena, but leaving no observable consequences 
through
nuclear physics effects. In doing so we are able to reconcile the claimed 
cosmological evidence for a changing fine structure constant with the 
tight constraints emerging from the Oklo natural nuclear reactor.
\end{abstract}


\maketitle

\section{Introduction}
Recently claimed observational evidence for a $\sim$
10ppm change of the fine structure constant 
over cosmological time between $z\simeq 0.5$ and $z=0$ 
brings an unexpected and interesting support to the
old idea of Dirac \cite{Dirac}. The result of Webb {\em et al.} \cite{Webb01}
(if it continues to defy alternative explanations by systematic effects 
\cite{Murphy:2002ve}) has serious implications for cosmology, 
astrophysics and 
particle physics, \cite{Sister}-\cite{Bek02}. 

The most powerful conclusion that follows from \cite{Webb01} 
is the existence of new fields,
massless or nearly massless, and different from gravitation, 
electromagnetism,
neutrinos and axions. Indeed, the ``promotion'' of a coupling constant 
$g$ into a function of time 
immediately implies that it should also depend on the rest of the coordinates, 
$g=g(x)$. The dependence of $g$ on $x$  in an interacting theory will 
inevitably generate a  kinetic term $\partial_\mu g \partial^\mu g$, so 
that  the ``coupling constant'' becomes a propagating field.
In order to avoid confusing terminology of ``changing alpha'', it is 
reasonable 
to introduce a new degree of freedom, a scalar field $\phi(x)$, via a 
chain of relations:  
\begin{eqnarray}
\fr{1}{g^2}F_{\mu\nu}F^{\mu\nu} &~\longrightarrow~&
\fr{1}{g^2(t)}F_{\mu\nu}F^{\mu\nu} ~\longrightarrow~ \nonumber \\
\fr{1}{g^2(x)}F_{\mu\nu}F^{\mu\nu} &~\longrightarrow~& 
\fr{1}{g^2_0}B_F(\phi(x))F_{\mu\nu}F^{\mu\nu} \,. \nonumber \\
\label{tx}
\end{eqnarray}
We postulate that the scalar field 
couples to the electromagnetic $F_{\mu\nu}F^{\mu\nu}$ via some unknown 
dimensionless function $B_F(\phi(x))$. Using the freedom in the
definition, one can always choose 
$B_F(\phi(z=0))=1$, so that $g_0$ is the coupling constant ``today'', i.e. at 
the redshift $ z = 0$. 

The dependence of gauge coupling constants on coordinates can hardly be 
regarded as a theoretical novelty. Indeed, in string theory/supergravity
all effective coupling constants depend on scalar field(s)
$\phi_i(x)$, 
so that 
$g=g(\phi_i(x))$. However, the {\em late} change of $g$, a few billion 
years after the Big 
Bang, implies that some of these fields have essentially near flat potentials, 
i.e. their masses must be comparable to the Hubble parameter at 
the present epoch $\sim 10^{-32}$ eV. This is in contrast with the 
phenomenologically developed  models of string 
theory/supergravity, where all moduli are believed to 
receive electroweak scale masses after  
supersymmetry breaking, which makes them completely frozen 
long before nucleosynthesis, not mentioning the present epoch. 
Therefore, such a late evolution of the effective $\alpha$ comes as 
a total surprise for particle physics.  

In general, a nearly massless scalar poses a serious 
problem of naturalness, which only worsens if this scalar is 
interacting \cite{BDD}. 
Indeed, the inclusion of a non-renormalizable interaction of the form 
$ B_F(\phi)F^2$ at the quantum level requires the existence of a 
momentum cut-off $\Lambda_{UV}$. Unfortunately,, any particle physics 
motivated choice of $\Lambda_{UV}$ destabilizes the quintessence 
potential, i.e. it could induce a mass term much 
larger than the required $m_\phi \sim H_0$. 
However, because the nature of this fine-tuning is similar to the 
one required for 
the smallness of the cosmological constant, we choose to proceed hoping that 
one day both problems could be resolved simultaneously. 

The study of the interaction (\ref{tx}) has a long history. Initially, it was 
examined by Bekenstein \cite{Bek}, who introduced the exponential form 
for the  
coupling of the scalar field to the electromagnetic Lagrangian which 
in practice can always be taken in the linear form $B_F(\phi(x))= 1-
\zeta \kappa \phi$, where $\kappa^2 = 8 \pi G$.
 He showed that the exchange by this scalar
leads to the effective non-universality of the gravitational force which
bounds the possible size of $\zeta$ at one per mil level. The cosmological 
evolution of 
$\phi$ is dynamical and it starts immediately after the transition to
a  matter-dominated 
Universe. In principle, the existence of a non-zero $\zeta$ alone
is sufficient to ensure the cosmological evolution of $\phi$, driven by the 
electromagnetic portion of the baryon mass density. However, in the 
minimal Bekenstein model where $\zeta$ is the only source for $\phi$, 
the resulting change of $\alpha$ is way too small to be interesting:
$|\alpha(z\sim 1) - \alpha_0|/\alpha_0 \la 10^{-10}-10^{-9}$ \cite{Bek,LS,OP}
An enhancement by up to five-six orders of magnitude may occur by abandoning 
the  minimalistic approach and introducing a ``generalized'' Bekenstein model 
where $\phi$ is driven either by $O(1)$ couplings to dark matter or 
by its own 
potential \cite{OP,SBM}, in which case $\Delta \alpha/\alpha$ 
may be consistent 
with the results of Refs.~\cite{Webb01}. 

Quite independently from the physics of changing couplings, there has 
been a tremendous increase of interest towards cosmological models 
with a light scalar field as a possible source of 
dark energy or quintessence.  
Evidence for the change of the effective coupling 
constant is a strong argument in favor of the 
existence of a scalar field with the wave length comparable to the size of 
the Universe which thus makes it a good candidate for quintessence.
Even though the concept of quintessence and the choice if its 
potential looks completely {\em ad hoc}, some quintessence models have an 
advantage over a ``simple'' explanation of dark energy by a non-zero 
cosmological constant. Indeed, certain forms of potentials allow for an 
attractor-type solution, in which case the late cosmological evolution 
of $\phi$ has no memory of the initial conditions 
that could be chosen (almost) arbitrarily.  The evolution 
of the energy density of the $\phi$ field is 
subdominant with respect to the energy density
of matter and radiation, and only during the last stages of the cosmological 
evolution does it become a dominant component. 
It is thus intriguing to identify the field $\phi$ responsible for the time 
dependence of $\alpha$ with tracking quintessence and study 
possible patterns for 
$\Delta\alpha/\alpha$ as a function of cosmological time.

In this paper, we present a thorough analysis of a large class of
 scalar field models with tracking behavior 
supplemented by a coupling to the electromagnetic field. A similar, but 
more restricted comparison has recently been made in \cite{AG}. Where 
appropriate we will compare our results to those. 

A successful
model must satisfy the following present constraints:
\begin{itemize}
\item[(1)] Oklo natural reactor $(z = 0.14)$ \cite{Oklo,DD,Fujii,OPQ}:
\begin{equation}
\label{oklobound}
\left| \frac{\Delta \alpha}{\alpha} \right| \la 10^{-7} \,;
\end{equation}

\item[(2)] Decay rate $^{187}{\rm Re} ~\rightarrow  ~^{187}{\rm Os}$, 
$(z = 0.45)$ \cite{OPQ}:
\begin{equation}
\label{decaybound}
\left| \frac{\Delta \alpha}{\alpha} \right| \la 
10^{-6} \,;
\end{equation}

\item[(3)] Cosmic microwave background radiation $(z = 10^{3})$
  \cite{Avelino:2001nr}:
\begin{equation}
\label{cmbbound}
\left| \frac{\Delta \alpha}{\alpha} \right| < 10^{-2} \,;
\end{equation}

\item[(4)] Big Bang nucleosynthesis $(z = 10^{8} - 10^{10})$ 
\cite{Ichikawa:2002bt}:
\begin{equation}
\label{nuclbound}
\left| \frac{\Delta \alpha}{\alpha}\right| \la 
2 \times 
10^{-2} \,;
\end{equation}

\item[(5)] Atomic clocks $(z = 0)$ \cite{Marion:2002iw}:
\begin{equation}
\label{atmbound}
\left| \frac{\dot{\alpha}}{\alpha}\right| < 4.2 \times 
10^{-15} ~{\rm yr}^{-1} \,,
\end{equation}
where a dot represents differentiation with respect to cosmic time;
\item[(6)] Equivalence principle tests \cite{OP}:
\begin{equation}
\label{zetabound}
|\zeta| < 10^{-3} \,.
\end{equation}
\end{itemize}

In addition, we take into account the claimed evidence for the variation of
the fine structure constant. We will assume 

\begin{itemize}
\item[(7)] Quasar absorption lines $(z = 3)$ \cite{Webb01}:
\begin{equation}
\label{webbvalue}
\frac{\Delta \alpha}{\alpha} = -10^{-5} \,.
\end{equation}
\end{itemize}
 
We first address a minimal Bekenstein-like coupling and find the
behavior of 
the fine structure constant as a function of the redshift for the 
exponential, power-like 
and exponential times power-like potentials. 

In all cases we 
analyze the scalar field evolution in the tracking regime i.e., 
when the energy
density of the scalar field remains sub dominant. In this case we are
able to present the {\em lower} bound on $\zeta$ consistent with 
tracking behavior and 
a non-zero change of the fine structure constant \cite{Webb01}. 
The upper limit in
this case is furnished by the direct experimental limit on the 
non-universality of the gravitational force. 
As a separate exercise, we consider a scalar field as quintessence, i.e.
when the tracking regime changes to the scalar field domination at $z\sim 1$. 
In this case, we are able to determine the allowed window for 
$\zeta$, without referring to the constraints from the 
non-universal gravitational
forces. 
Also in the context of a quintessence model, we consider the case of 
the tracker field driven by the combination of the self potential 
and the coupling to dark matter.

We confront the predictions for 
$\Delta \alpha/\alpha$ with the above limits from various 
astronomical and terrestrial observations, and 
find that the present day limits on $\dot{\alpha}/\alpha$ 
and constraints from the Big Bang nucleosynthesis are generally safe. 
On the other hand, the  $\Delta \alpha/\alpha $ coming from the 
Oklo natural reactor 
\cite{Oklo,DD,Fujii,OPQ} and from the meteoritic abundance data \cite{OPQ}
are generally inconsistent with tracking behavior.
In the case of a  non-minimal coupling of quintessence to the 
electromagnetic field, the results lose their predictivity as almost any 
profile for $\alpha(t)$ can be achieved by an appropriate 
choice of $B_F(\phi)$. 

We go on to discuss a novel possibility that the effective change of 
the coupling 
constant occurs only in atomic interactions  while $\Delta \alpha/\alpha$ 
remains effectively zero for nuclear physics phenomena. 
The construction we propose is based on the idea that the coupling 
between photons and 
the scalar field may change for off-shell photons
and has a form factor at some energy above the electron mass. 
This leads to the suppression 
of the coupling between the scalar field and the nuclear 
levels giving a chance to reconcile  
the results of Webb {\em et al.} and the bounds coming from 
Oklo phenomenon without fine-tuning. 

This paper is organized as follows: in the next section we analyze in detail 
four types of tracking scalar fields and determine the allowed
values for $\zeta$. 
Section 3 addresses the issue of the non-minimal 
$B_F(\phi)$ and 
introduces the model where $\zeta$ has a form factor. We reach our conclusions 
is section 4.

\section{Minimal coupling of quintessence to the electromagnetic field}

We study a class of models of a neutral scalar field coupled to 
electromagnetism:
\begin{eqnarray}
S = &-& \frac{1}{2 \kappa^2} \int d^4x \sqrt{-g}~ R \nonumber \\
    &+&
   \int d^4x \sqrt{-g} ~({\cal L}_{\phi} + {\cal L}_B + {\cal L}_{\phi F}) \,,
\end{eqnarray}
where $R$ is the scalar curvature. In this paper, we deal with the  
cosmological 
evolution and take the stress-energy tensor of
the matter fields in a homogeneous form corresponding to 
a background perfect fluid with the energy-momentum tensor given by
\begin{eqnarray}
{T_B}_{\mu\nu} &\equiv& \frac{2}{\sqrt{-g}}\frac{\delta(\sqrt{-g} {\cal
L}_B)}{\delta g^{\mu\nu}} \nonumber \\
&=& (\rho_B^{}+p_B^{})u^{\mu} u^{\nu} - g_{\mu\nu}p_B^{} \,. 
\end{eqnarray}
Throughout the paper, we employ $(+,-,-,-,)$ signature for the metric tensor.
The pressure $p_B^{}$ and energy density $\rho_B^{}$ are related by
the equation of state $p_B^{} = (\gamma-1) \rho_B^{}$.
$\gamma$ is taken to be a constant (e.g. $\gamma = 4/3$ for
radiation and $\gamma = 1$ for matter).
The Lagrangian density for the scalar field $\phi$ is
\begin{equation}
{\cal L}_{\phi} = \frac{1}{2} \partial^{\mu}\phi\partial_{\mu}\phi
- V(\phi) \,.
\end{equation}
The associated energy-momentum tensor of the scalar field then reads
\begin{equation}
{T_{\phi}}_{\mu\nu} = \partial_{\mu}\phi\partial_{\nu}\phi -
g_{\mu\nu} \left( \frac{1}{2} \partial^{\mu}\phi\partial_{\mu}\phi
-V(\phi) \right) \,.
\end{equation}
The interaction term between the scalar field and the electromagnetic
field is
\begin{equation}
{\cal L}_{\phi F} = - \frac{1}{4} B_F(\phi) F_{\mu\nu}F^{\mu\nu},
\label{bc}
\end{equation}
where $B_F(\phi)$ allows for the evolution in $\phi$.  In this paper we 
 adopt a linear dependence on $\phi$ such that 
$B_F(\phi) = 1-\zeta\kappa (\phi-\phi_0)$, where the subscript 0
represents the present value of the quantity.
The effective fine structure constant depends on the value of $\phi$ as
$\alpha = \alpha_0/B_F(\phi)$.  Therefore, we have 
\begin{equation}
\label{defDelta}
\frac{\Delta \alpha}{\alpha} \equiv \frac{\alpha-\alpha_0}{\alpha_0} =
\zeta \kappa (\phi - \phi_0) \,.
\end{equation}
We will consider a spatially flat Friedmann-Robertson-Walker (FRW)
Universe with metric $ds^2 = dt^2 - a^2(t) d{\bf x}^2$. We further
assume a nearly homogeneous scalar field so that the spatial
derivatives of the field can be neglected. Hence, we can define
analogously to the case of
a perfect fluid, the energy density and pressure of a
scalar field as, $\rho_{\phi}^{} = \dot{\phi}^2/2 + V$
and $p_{\phi}^{} = \dot{\phi}^2/2 - V$, respectively.

The governing equations of motion are
\begin{eqnarray}
\dot{H} &=&- \frac{\kappa^2}{2} \left( \gamma \rho_B^{} + \dot{\phi}^2
\right) \,, \\
\label{eqcontinuity}
\dot{\rho}_B^{} &=& -3 \gamma H \rho_B^{} \,, \\
\label{eqscalar}
\ddot{\phi} &=& - 3 H \dot{\phi} - \frac{d V}{d \phi}  \,,
\end{eqnarray}
subject to the Friedmann constraint
\begin{eqnarray}
H^2 = \frac{\kappa^2}{3} \left(\rho_B^{} + \rho_{\phi}^{} \right) \,.
\end{eqnarray}
The term ${\cal L}_{\phi F}$ can be neglected in the equations 
determining the dynamics of $\phi$
as its average is zero for photons, and also quite small,
$O(10^{-3})$ for baryons \cite{Bek,OP}. A notable exception is when 
$\phi$ couples strongly and directly to dark matter \cite{OP}.

In the following subsections we will consider different forms of
the scalar potential $V(\phi)$ and analyze 
two possible cases of evolution of the field: (i)
slow roll down the potential $V(\phi)$ controlled by the friction
provided by the background fluid (tracker evolution); (ii) 
the field provides the source of dark energy, and 
is just starting to dominate the 
dynamics leading the universe into a period of accelerated expansion.
We will start by estimating the value of the coupling $\zeta$
consistent with Eq.~(\ref{webbvalue}) by setting
\begin{equation}
\label{zetavalue}
\zeta = - \frac{10^{-5}}{\kappa \phi(z=3) - \kappa \phi(z=0)} \,.
\end{equation}
One should bare in mind that we are not attempting to fit the data to
specific scalar potentials but instead to extract an estimate of
$\zeta$ consistent with the data.

We calculate the ratio of the variation of $\alpha$ with time at
present using the relation
\begin{equation}
\label{dotalpha}
\frac{\dot{\alpha}}{\alpha} = - \zeta \frac{d (\kappa \phi)}{d (1+z)}~ H_0
\,,
\end{equation}
where $H_0$ is the value of the Hubble constant today, 
$H_0 = h/9.78~ \times 10^{-9}~{\rm yr}^{-1}$. We will use $h = 0.65$
throughout the paper.

\subsection{Exponential potentials}
\label{secexp}
For a pure exponential potential 
$V(\phi) = V_0 \exp(-\lambda \kappa \phi)$,
there exist just two possible
late time attractor solutions with quite different
properties, depending on the values of $\lambda$ and
the background's equation of state $\gamma$ \cite{Copeland:1997et}:

(1) $\lambda^2 > 3\gamma$. The late time attractor
 is one where
the scalar field mimics the evolution of the barotropic fluid
 with $\gamma_{\phi} = \gamma$, and the relation
$\Omega_{\phi} = 3\gamma/\lambda^2$ holds.

(2) $\lambda^2 < 3\gamma$. The late time attractor is the
scalar field dominated
solution ($\Omega_{\phi} =1$) with $\gamma_{\phi} = \lambda^2/3$.

The second  regime can be discarded as unrealistic.
Indeed, there is a tight constraint on the allowed magnitude of  
$\Omega_{\phi}$ at the time of Big Bang nucleosynthesis, 
$\Omega_{\phi}(1~{\rm MeV}) < 0.045$
\cite{Bean:2001wt}. Moreover, we must allow time for the formation of structure
before the universe starts accelerating, which requires $\Omega_{\phi}$
to be subdominant after the CMB decoupling. Therefore, the transition to 
$\Omega_{\phi} \simeq 1$ should occur at the present epoch.
For this model, this is possible only at the expense of extreme fine tuning
of the initial value $\rho_{\phi}^{\rm in}$, a type of fine-tuning 
that goes against the whole idea of the tracker/quintessence field. 
The first regime is perfectly compatible with all experimental data, 
{\em if} there is an additional component to dark energy which ensures 
the late time acceleration, as the $\phi$ field itself in this model 
does not allow for an accelerating expansion.

For now we simply assume that the scalar field responsible for the change
in the value of $\alpha$ is tracking and is subdominant today,
i.e. $\Omega_{\phi} < 1/2$. This imposes
a lower bound $|\lambda| \geq \sqrt{6}$. 
The evolution of the field in the
tracker regime is well known and is given by \cite{Ng:2001hs}
\begin{equation}
\label{expevol}
\kappa (\phi-\phi_0) = - \frac{3 \gamma}{\lambda} \ln (1+z) \,.
\end{equation}
Therefore, combining
Eqs.~(\ref{zetavalue}) and (\ref{expevol}), we obtain
\begin{equation}
\label{ztexp}
\zeta = \frac{10^{-5}}{\ln(4)} ~\frac{\lambda}{3 \gamma} \,.
\end{equation}
As we discussed before, if $\rho_\phi$ is subdominant today, then there must be
some other source of dark energy responsible for the accelerating
universe. The presence of a cosmological constant introduces an error of
less than $30 \%$ in the above estimate.
With $|\lambda|$ being limited from below, this imposes the lower bound on 
the possible size of the coupling, $\zeta > 6 \times 10^{-6}$.
On the other hand, $\zeta$ is limited from tests of the equivalence 
principle, $|\zeta| < 10^{-3}$, which  we can use to further constrain  
the value of the
parameter $\lambda$ in a matter dominated universe:
\begin{equation}
\sqrt{6} < |\lambda| < 415 \,.
\end{equation}

From Eq.~(\ref{dotalpha}) we immediately extract that for this
potential $\dot{\alpha}/\alpha = 4.8 \times 10^{-16}~ {\rm yr}^{-1}$
consistent with the atomic clocks bound Eq.~(\ref{atmbound}).

It is also instructive to check whether this range of couplings is 
consistent with 
constraints on $\Delta \alpha/\alpha$ coming from the Big Bang
nucleosynthesis. In the radiation dominated era, the evolution of the
scalar field is given by
\begin{equation}
\label{exprad}
\kappa (\phi-\phi_0) = - \frac{3}{\lambda} \ln (1+z_{\rm eq}) 
- \frac{4}{\lambda} \ln \left( \frac{1+z}{1+z_{\rm eq}} \right) \,,
\end{equation} 
where $1+z_{\rm eq}$ is the redshift at which the energy density of the
radiation and matter components have the same value, $z_{\rm eq} =
\Omega_{\rm M}^0/\Omega_{\rm R}^0 \approx 5 \times 10^{3}$. 
Substituting into Eq.~(\ref{defDelta}) for the estimated value of $\zeta$,
 it is easy to verify that  
$\Delta \alpha/\alpha = -2 \times 10^{-4}$ at redshift $ z = 10^{10}$, which
is fully consistent with the bound in Eq.~(\ref{nuclbound}).

\subsection{Power law potentials}
For an inverse power law potential of the form $V(\phi) =
M^{4+n}/\phi^n$ the cosmological evolution of the field is 
given by \cite{Ng:2001hs}
\begin{equation}
\label{powevol}
\kappa (\phi -\phi_0) = \kappa \phi_0 
\left[ -1 + (1+z)^{-3 \gamma/(2+n)} \right] \,,
\end{equation}
where the value of $\phi$ at $z=0$ is given by 
\begin{equation}
\label{phi0pow}
\phi_0 = \left[\frac{3}{2} \frac{\kappa^2 \rho_B^0}{M^{4+n}}
\frac{1}{n^2}\frac{\gamma n}{n+2}
\left(2-\frac{\gamma n}{n+2}\right)\right]^{-1/(2+n)} \,.
\end{equation}
Combining Eq.~(\ref{powevol}) with Eq.~(\ref{zetavalue}) we can
estimate the coupling with electromagnetism through
\begin{equation}
\zeta = -\frac{10^{-5}}{-1 + 4^{\beta}}~ (\kappa \phi_0)^{-1} \,,
\end{equation}
where $\beta = -3 \gamma/(2+n)$.

\begin{figure}[!ht]
\centerline{
\includegraphics[width=8cm]{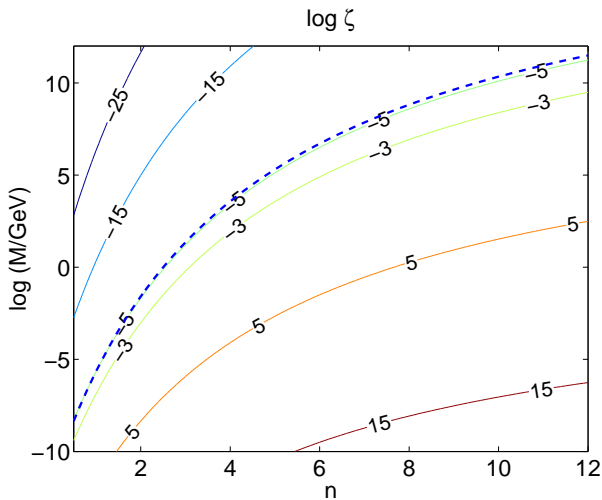}
}
\caption{Contour plot for $\log \zeta$ with respect to the possible choices
of $n$ and $M$ in the inverse power law potential. The dashed line
represents the maximum value for $M$ that allows tracking evolution of
the field at present.}
\label{figure3}
\end{figure}

In Fig.~\ref{figure3} we show a contour plot 
for $\log \zeta$ 
as a function of the parameters $n$ and $M$ in the scalar potential 
consistent with $\Delta\alpha/\alpha$ suggested by the QSO data. 
A few comments are in order at this point. Note that at fixed $n$, $\zeta$ 
falls off with increasing $M$. This behavior follows from  
Eq.~(\ref{phi0pow}): as the mass scale $M$
increases so does $\phi_0$ and consequently $\kappa(\phi-\phi_0)$. 
The dashed line on this plot 
represents the upper
bound on the mass scale $M$ coming from our initial assumption that
the field is tracking today. For a given $n$, the value of this
upper bound corresponds to the case when the field has just begun to
dominate. This is the case of quintessence. 
Moreover, a value larger than this means that the scalar
field has been dominating the evolution of the universe for a long
time preventing any formation of structure. Thus everything above the 
dashed line is excluded. 
The equivalence principle bound, $|\zeta| < 10^{-3}$, and the upper
bound on $M$ produce an allowed band for the pair $(n,M)$ limited by 
the contours $-5$ and $-3$ in Fig.~\ref{figure3}.  

\begin{figure}[!ht]
\centerline{
\includegraphics[width=8cm]{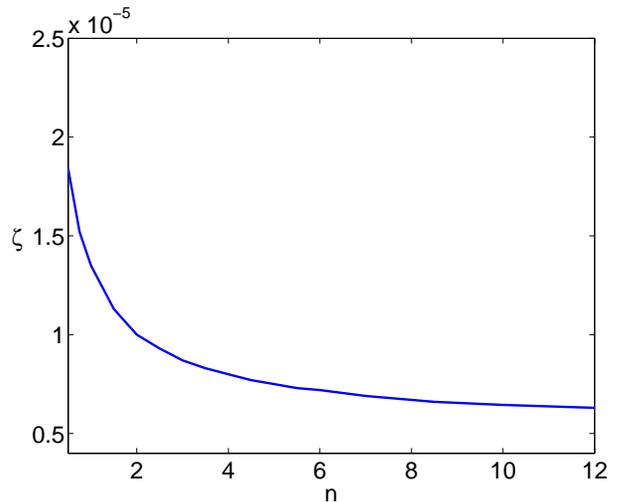} }
\caption{Variation of $\zeta$ with the parameter $n$ for an inverse
  power law quintessence model.}
\label{figure3.1}
\end{figure}

The case of quintessence (i.e. along the dashed line in Fig.~\ref{figure3}) 
is more involved and requires a special approach as the field today
is no longer in an
attractor regime but in the transition between the tracker and a
scalar field dominated solution. Nevertheless, from Fig.~\ref{figure3} 
we expect the value of the coupling to be close to $\zeta \approx 10^{-5}$.
In Fig.~\ref{figure3.1} we show the 
variation of parameter $\zeta$ with $n$ for a Universe with $\Omega_{\phi} =
0.7$ today. This curve was obtained by solving numerically the
evolution equations of the field followed by extraction of the
coupling $\zeta$ through Eq.~(\ref{zetavalue}). 
One should point out that models with $n > 3$ lead
to an equation of state for the quintessence field in clear
disagreement with observations as the latter indicate $\gamma_{\phi}
\equiv p_{\phi}/\rho_{\phi} > 0.4$ \cite{Efstathiou:1999tm}. 
On the other hand, models with $n < 1$ do not
alleviate the initial conditions problem, which was the great advantage
of tracking quintessence models over the bare cosmological constant.
 Thus, the range of interesting
$n$, $1\le n\le 3$, in the case of quintessence defines the window of the 
coupling constant $\zeta$ that agrees with QSO evidence for changing alpha:
\begin{equation}
 8.6 \times 10^{-6} < \zeta < 1.4 \times 10^{-5} \,.
\end{equation}

\begin{figure}[!ht]
\centerline{
\includegraphics[width=8cm]{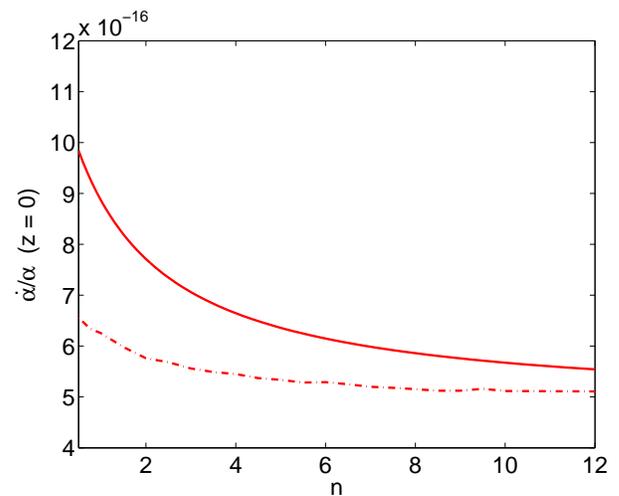} }
\caption{$\dot{\alpha}/\alpha$ at present 
for a tracker evolution along an inverse
  power law potential (solid line) and for a quintessence model
  (dot-dashed line).}
\label{figure4}
\end{figure}

We can calculate now the value of $\dot{\alpha}/\alpha$ at present
for a given $n$ using Eq.~(\ref{dotalpha}). In Fig.~\ref{figure4} we
show the shapes of these curves for the tracker and quintessence
cases. 
 
Finally, as in the case of the exponential potential, the 
BBN bounds on $\Delta \alpha$ are satisfied with this model.
The evolution of the field in the radiation dominated epoch is
\begin{eqnarray}
\kappa (\phi-\phi_0) = \kappa \phi_0 
\left[ -1 + (1+z_{\rm eq})^{-3/(2+n)} ~\times \right. \nonumber \\ 
\left. \left(\frac{1+z}{1+z_{\rm eq}}\right)^{-4/(2+n)} \right] \,,
\end{eqnarray}
hence, we can verify that \newline
$10^{-5} < |\Delta \alpha/ \alpha (z = 10^{10})| <  4 \times
10^{-5}$, well within the bound of Eq.~(\ref{nuclbound}).

\subsection{SUGRA type potentials}
Let us now consider the more general potential studied in Ref. 
\cite{Ng:2001hs} which we write in the form
\begin{equation}
V(\phi) = V_0 e^{\lambda (\kappa \phi)^{\beta}} (\kappa \phi)^{n} \,.
\label{genpot}
\end{equation}
See Refs. \cite{Brax:1999gp,Copeland:2000vh} that motivate this form within a 
supergravity context, hence the name adopted here.
For negative $n$ this potential has a minimum
with non-vanishing energy density which upon convenient choice of the
energy scale $V_0$ can explain the present acceleration of the universe.
The evolution of the field cannot be written in an explicit form but,
for $(\kappa \phi)^{\beta} \gg n/\lambda
\beta$ the evolution in the tracker regime can be written in terms of a 
perturbative solution to order $k$ written recursively,  such that
\begin{eqnarray}
\label{sol1}
\lambda (\kappa \phi)^{\beta}_{(1)} &=& \Phi_i  + 3 \gamma \ln(1+z) 
\nonumber \,, \\
\lambda (\kappa \phi)^{\beta}_{(k)} &=&  \lambda (\kappa
\phi)^{\beta}_{(1)} + \frac{2-n}{\beta} \ln(\kappa
\phi)^{\beta}_{(k-1)} 
\nonumber \\
&~& - 2 \ln \left[\lambda \beta
                     (\kappa \phi)^{\beta}_{(k-1)}+ n \right]   \,
\end{eqnarray}
where 
\begin{equation}
\Phi_i = \ln \left[\frac{3}{2} \frac{\rho_B^0}{V_0} \gamma(2-\gamma)
  \right] \,.
\end{equation}
and the last logarithm term only exists if $\lambda \neq 0$.
The case $(\kappa \phi)^{\beta} \ll n/\lambda
\beta$ is simply the power law case that we studied in part B.

\begin{figure}[!ht]
\centerline{
\includegraphics[width=8cm]{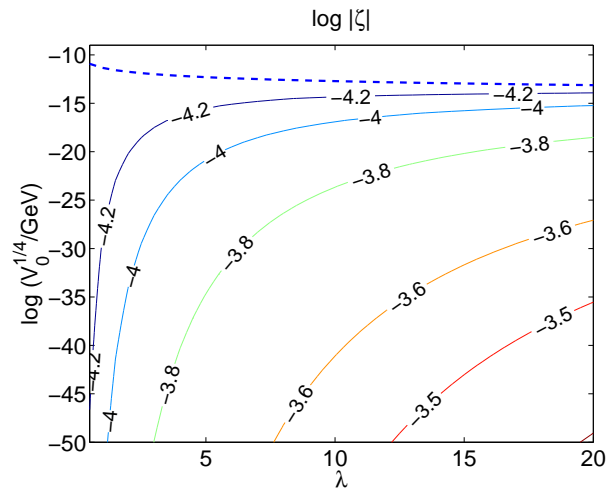}
}
\caption{ Contour plot for $\log |\zeta|$ with respect to the possible
choices of $\lambda$ and $V_0$ for the model of Eq.~(\ref{genpot}) with 
$n = -11$ and $\beta = 2$.}
\label{figure51}
\end{figure}

Following the method adopted above to estimate $\zeta$, its value is
given by
\begin{equation}
\zeta = - \frac{10^{-5}}{\kappa \phi_{(k)}(z=3)- \kappa \phi_{(k)}(z=0)} \,,
\end{equation}
where $\phi_{(k)}$ is given by Eq.~(\ref{sol1}). In Fig.~\ref{figure51} we
show a contour plot for the value of $\zeta$ necessary to account for
the QSO data, using the second order approximation $k = 2$. 
In the tracker regime we expect a weak
dependence of any derived quantities on the parameter $n$. This
argument follows from the initial assumption that the field is rolling
along the exponential side of the potential.
We have fixed $\beta = 2$, $ n= -11$ and made $\lambda$ and $V_0$ free 
parameters. The particular case $\lambda = 1/2$ 
leads to the form of the potential in
Ref.~\cite{Brax:1999gp}.
In Fig.~\ref{figure51} we show a contour plot 
for $\log \zeta$ with respect to the possible
choices of $n$ and $V_0$.
The dashed line represents the value of $V_0$ that corresponds to the
present energy density of dark energy assuming that the field has
already reached the minimum of the potential at $z = 0$.
This method of estimating $V_0$ is particulary good for
$\lambda > 1$.
%

As we have seen in the inverse power law potential case, as we approach the
dashed line (or quintessence like behavior) the value of the coupling
approaches $\zeta \approx 10^{-5}$. For very small values of $V_0$ and
large $\lambda$ we enter the region of parameters disfavored by the
equivalence principle experiments. Again, as in the inverse power law
potential, we obtain an allowed band, now for the pair $(\lambda,V_0)$, 
limited by the contours $-5$ and $-3$. For increasing $\beta$ one
can verify that this band becomes narrower. 

\begin{figure}[!ht]
\centerline{
\includegraphics[width=8cm]{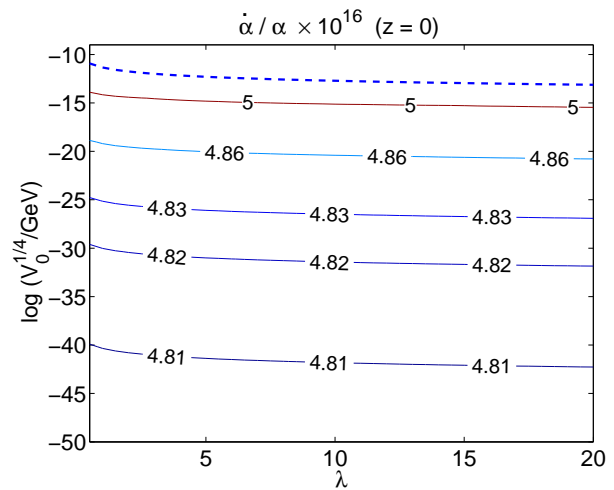}
}
\caption{ Contour plot for $\dot{\alpha}/\alpha$ with respect to the possible
choices of $\lambda$ and $V_0$ for the model of Eq.~(\ref{genpot}) with 
$n=-11$ and $\beta = 2$.}
\label{figure61}
\end{figure}

\begin{figure}[!ht]
\centerline{
\includegraphics[width=8cm]{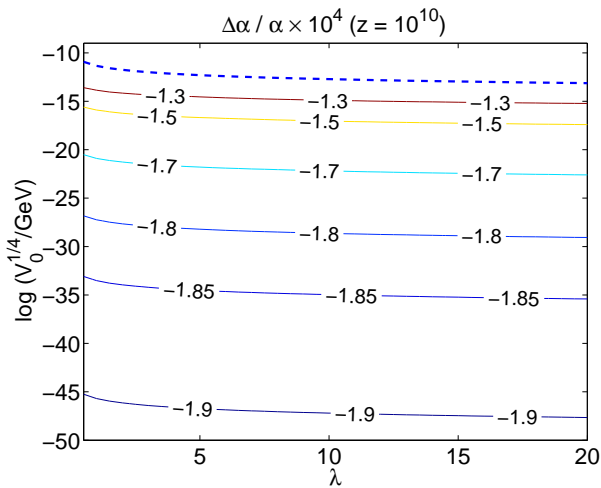}
}
\caption{ Contour plot for $\Delta\alpha/\alpha$ at nucleosynthesis 
with respect to the possible
choices of $\lambda$ and $V_0$ for the model of Eq.~(\ref{genpot}) with 
$n=-11$ and $\beta = 2$.}
\label{figure71}
\end{figure}

To calculate the variation of $\alpha$ at the time of nucleosynthesis
we just have to realize that the only modification in Eq.~(\ref{sol1})
is that in the radiation era
\begin{equation}
\lambda (\kappa \phi)^{\beta}_{(1)} = \Phi_i  + 
3 \ln(1+z_{\rm eq}) + 4 \ln \left(\frac{1+z}{1+z_{\rm
    eq}}\right) \,.
\end{equation}
In Figs.~\ref{figure61} and \ref{figure71} 
we show that both the values of $\dot{\alpha}/\alpha$ at present and
$\Delta \alpha /\alpha$ at nucleosynthesis 
for this model are within the current experimental and observational
bounds. A striking property is that these quantities are extremely
insensitive to the values chosen for the parameters of the model, a
feature we came across before in the previous models. This should not
come as a surprise because when we estimate $\zeta$ by fixing $\Delta
\alpha/\alpha = 10^{-5}$ at $z=3$, we are effectively demanding that
the late time evolution is roughly the same for any choice of parameters.

%

\subsection{Coupled quintessence}  

We now consider a linear coupling between the scalar field
and the background fluid such that 
Eqns.~(\ref{eqcontinuity}) and (\ref{eqscalar}) are now written in the
form:
\begin{eqnarray}
\label{qeqom2}
\dot{\rho}_B^{} + 3H \gamma \rho_B^{} &=& 
-\kappa C \dot{\phi} (\rho_B^{}-3p_B)  \,, \\
\label{qeqom1}
\ddot{\phi} + 3H\dot{\phi} + \frac{d V}{d \phi} &=& \kappa 
C(\rho_B^{} - 3p_B) \,.
\end{eqnarray}
Note that the background energy density and pressure enter in the 
combination $\rho_B-3p_B$ to account for the fact that
the background radiation does not provide any source for $\phi$. 

We assume an exponential potential for the scalar field
$V(\phi) = V_0 \exp( -\lambda \kappa \phi )$. 
Such a coupling can arise after a conformal transformation to the
Einstein frame in Brans-Dicke theory \cite{Holden:1999hm}.
We note, however, that our model cannot be considered as pure 
Brans-Dicke theory as the latter does not provide a 
$\phi F_{\mu\nu}F^{\mu\nu}$ coupling. 

It is known that the system has two cosmologically interesting
attractor solutions \cite{Amendola:1999qq,Holden:1999hm,Amendola:1999er}:

(1) a scaling solution with
\begin{eqnarray}
\label{omegaphi}
\Omega_{\phi} &=& \frac{C^2 - C \lambda + 3\gamma}{(C-\lambda)^2} \,,
\\
\gamma_{\phi} &=& \frac{3\gamma^2}{C^2 - C\lambda + 3\gamma} \,;
\end{eqnarray}

(2) a scalar field dominated solution with $\Omega_{\phi} = 1$ and
$\gamma_{\phi} = \lambda^2/3$,
identical to the minimally coupled case (see Sec.~\ref{secexp}).

For solution (1), the power law expansion $a \propto t^p$ is given by
\begin{equation}
p = \frac{2}{3 \gamma} \left(1-\frac{C}{\lambda} \right) \,,
\end{equation}
(not to be confused with pressure). The solution is inflationary for $p >
1$, that is, for $-\lambda < 2 C$ for a matter background. 

The evolution of the field for this kind of coupling is
\begin{equation}
\kappa (\phi-\phi_0) = - \frac{3 \gamma}{\lambda-C} \ln (1+z) \,,
\end{equation}
which reduces to Eq.~(\ref{expevol}) for $C = 0$.
Therefore, following the same approach as before, we have that in this
case the coupling of the scalar field
to the electromagnetic Lagrangian is
\begin{equation}
\zeta = \frac{10^{-5}}{\ln(4)} ~\frac{\lambda-C}{3 \gamma} \,,
\end{equation}
in order to explain the Webb {\it et al.} variation of $\alpha$.

\begin{figure}[!ht]
\centerline{
\includegraphics[width=8cm]{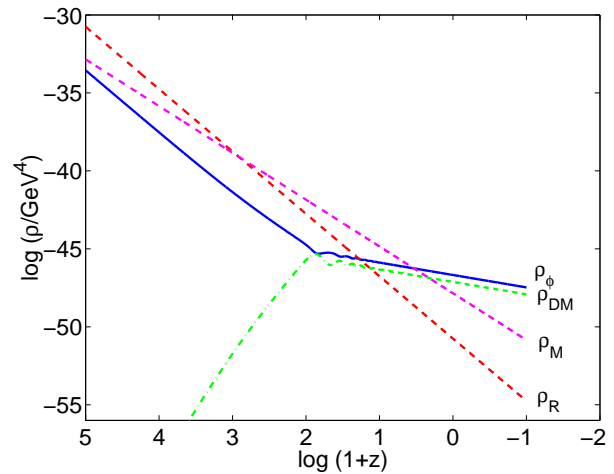} }
\caption{\label{quinte4} The late time evolution of the energy
densities of the scalar field $\rho_{\phi}^{}$, dark matter
$\rho_{\rm DM}^{}$, baryons $\rho_M^{}$ and radiation
$\rho_{\gamma}^{}$, in
the coupled quintessence model when all of the dark matter is coupled
to the scalar field.}
\end{figure}

Let us consider the scenario where the scalar field is coupled primarily to 
the dark matter and its coupling to baryons is mediated only by photons via
$\phi F_{\mu\nu}F^{\mu\nu}$ interaction to suppress the violation of 
equivalence principle.    
In Fig.~\ref{quinte4} we show the late time evolution of the energy
densities of the scalar field, dark matter, baryons and radiation in
the case when the scalar field is coupled to all of the dark matter.
%
%
We see that at early times the field scales like
radiation (as $p_{\rm rad} = \rho_{\rm rad}/3$ and hence the right
hand side of Eqs.~(\ref{qeqom2}) and (\ref{qeqom1}) vanishes), and
it scales like ordinary matter after the radiation-matter
transition. At this stage one can take $C$ to be effectively zero in
Eqs.~(\ref{qeqom2}) and (\ref{qeqom1}) following the assumption 
that the $\phi$ couples to baryons very weakly.
Only when the dark matter
contribution becomes important, does the scalar field reach the
attractor solution (1) above, driving the universe to an accelerated
expansion with $\gamma_{\phi}$ very close to zero, in good agreement
with observations.
However, the dark matter should play an important role in the process
of structure formation, which would  not be possible if it becomes dominant at 
such a late stage. For this scenario to be viable we are forced to introduce
a component of dark matter that is coupled to the scalar field and a
component that is not. An alternative approach, 
Ref.~\cite{Amendola:2000uh}, would be to postulate a non-linear 
coupling $C = C(\phi)$. 

\begin{figure}[!ht]
\centerline{
\includegraphics[width=8cm]{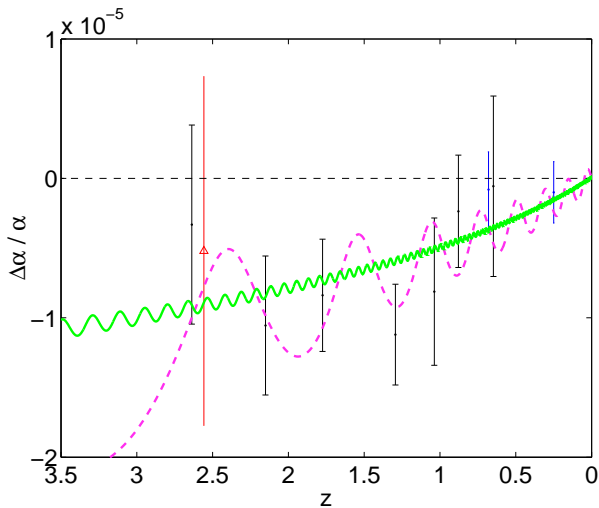} }
\caption{\label{coupl} Evolution of $\Delta \alpha/
\alpha$ for a coupled quintessence model against the observational QSO
data. The coupled dark matter component has a contribution 
$\Omega_{c{\rm DM}} = 0.05$ and $\Omega_{\phi} = 0.7$ today. 
The solid curve corresponds to $\lambda = -100$ and the
dashed curve to $\lambda = -10$.}
\end{figure}

\begin{figure}[!ht]
\centerline{
\includegraphics[width=8cm]{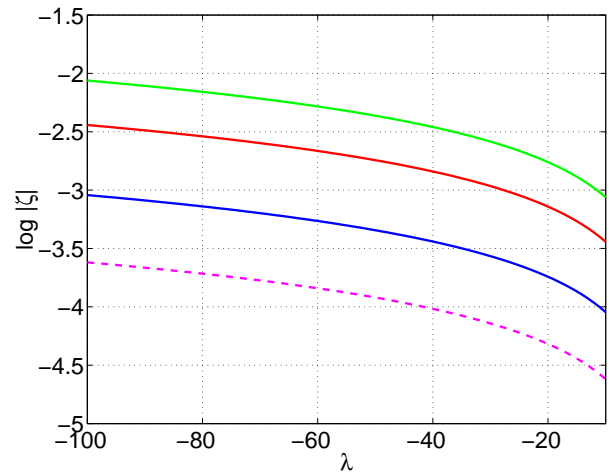} }
\caption{\label{omegadm} Dependence of $\zeta$ on the parameter
  $\lambda$ in the coupled quintessence model. The curves correspond to
different contributions of the coupled dark. 
From top to bottom we have $\Omega_{c{\rm DM}} =
  0.02, ~0.05, ~0.25$, and we assume $\Omega_{\phi} = 0.7$ today. The dashed
  curve corresponds to the variation of $\zeta$ in the uncoupled case of
  Sec.~\ref{secexp}.}
\end{figure}

In Fig.~\ref{coupl} we show the evolution of $\Delta \alpha/
\alpha$ for two different values of the slope of the potential
$\lambda$. The damped oscillations in the direction of small redshift 
reflect the fact that the solution is approaching the attractor.
In Fig.~\ref{omegadm} we show the dependence of the coupling
$\zeta$ on the slope of the potential $\lambda$ assuming different
present contributions of the dark matter component coupled to
the scalar field. The value of $C$ can be extracted from Eq.~(\ref{omegaphi}) 
where now we also have to take into account bi-component dark matter. Therefore, 
\begin{equation}
C = \lambda - \frac{\lambda - \sqrt{\lambda^2 - 12 \tilde{\Omega}_{c{\rm DM}}}}
{2 \tilde{\Omega}_{c{\rm DM}}} \,,
\end{equation}
is a good estimate for the value of the coupling in the tracker regime.
Here $\tilde{\Omega}_{c{\rm DM}} = \Omega_{c \rm DM}/(\Omega_{c \rm
  DM}+ \Omega_{\phi})$ and 
 $\Omega_{c{\rm DM}}$ stands for the density parameter of the dark matter 
 component coupled to the scalar field, 

We assume that $\Omega_{\phi} \approx 0.7$ today. Realistic
models of coupled quintessence must have a small $\Omega_{c{\rm DM}}$
for successful structure formation. Consequently, we
see from Fig.~\ref{omegadm}, that the equivalence principle
 constraints can only be satisfied
for larger values of $\lambda$ as the amount of $\phi$-independent component of dark matter
increases. For large values of $\lambda$ we have
seen from Fig.~\ref{coupl} that the attractor is reached at a later stage,
hence the field and necessarily $\Delta \alpha/ \alpha$ oscillate
heavily at low redshifts. 

As in the case of a pure exponential potential it turns out that
$\dot{\alpha}/\alpha = 4.8 \times 10^{-16} ~{\rm yr}^{-1}$. However
note that this is the value that corresponds to the attractor solution. 
Since the field is oscillating around the latter, the value of 
$\dot{\alpha}/\alpha$ can
actually be larger by an order of magnitude or smaller, depending on 
the phase of oscillations.

The calculation of $\alpha$ at the time of primordial nucleosynthesis is more 
involved as it requires the calculation of the redshift where the
attractor behaviour
is reached. Our estimate for the relative change of $\alpha$ at BBN time, 
$\Delta \alpha/ \alpha \approx 10^{-3}$, is consistent with limit 
(\ref{nuclbound}). 
 
\subsection{Other quintessence models}
\begin{figure}[!ht]
\centerline{
\includegraphics[width=8cm]{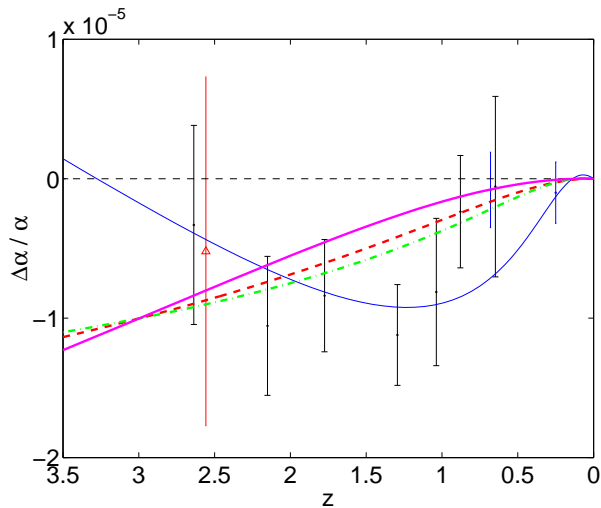} }
\caption{\label{figure10} Late time evolution of $\Delta 
\alpha/\alpha$ for the models in Table \ref{tabela} that pass the
Oklo bound: 2EXP model (c) in dashed line; A-S models (f) and (g) 
of Ref.~\cite{AG} in light solid and solid lines respectively;
SUGRA model (k) in dot-dashed line.}
\end{figure}

In this subsection we analyze some other 
quintessence
models suggested in the literature and evaluate if
a coupling to the electromagnetic field can provide a
realistic scenario compatible with terrestrial and cosmological
observations. In Fig.~\ref{figure10} we plot the late time evolution 
of $\Delta \alpha/\alpha$ for three models, the two exponential 
case \cite{Barreiro:1999zs}, the  Albrecht-Skordis 
model \cite{Albrecht:1999rm}, 
and variations to the SUGRA-inspired model proposed in \cite{Brax:1999gp}. 
Also included 
in the plot are the data points of Webb et al \cite{Webb01}. 
In Table \ref{tabela} we collect other relevant pieces of information
such as  typical values of the
quantities $\zeta$, $\dot{\alpha}/\alpha$ and $\Delta \alpha/ \alpha$
at nucleosynthesis. 
In addition, we indicate which models exhibit oscillatory behaviour at 
late times.  

\begin{table*}
\begin{center}
\begin{tabular}{|c|c|c|c|c|c|c|c|}
\hline
{\bf Model} & {\bf $A$} & {\bf $B$ } & 
{\bf $ \zeta \times 10^{5}$} &
{\bf $\dot{\alpha}/\alpha \times 10^{16}$} & 
{\bf $\Delta \alpha/\alpha (z = 0.14)$} &
{\bf $\Delta \alpha/\alpha (z = 10^{10})$} & {\bf osc} \\ 
\hline
2EXP {\hfill a} & 10 &  --8 & --20 & 8.6 & $-3.2 \times 10^{-6}$ &
$1.6 \times 10^{-3}$  & $\sqrt{}$ \\
{\hfill b} & 15 &  0.1 & 3.9 & 4.0 & $ -8 \times 10^{-7}$ & 
$-2.2 \times 10^{-4}$ & $\times$ \\
{\hfill c} & 10 & --0.5 & 3.9 & --0.2 & $-7 \times 10^{-8}$ & 
$-3.2 \times 10^{-4}$ & $\sqrt{}$ \\
\hline
A-S {\hfill d} & 10 & $0.9/A^2$ & --50 & --1.7 & 
$2 \times 10^{-6}$ &
$3.5\times 10^{-3}$  & $\sqrt{}$ \\
{\hfill e} & 6  & $1.1/A^2$ & 4.5 & 1.2 & 
$-2 \times 10^{-6}$ &
$-5.5 \times 10^{-4}$  &  $\times$ \\
{\hfill f} & 6 & $0.985/A^2$ & 11.2 & --1.4 & $-1 \times 10^{-8}$ &
$-1.4 \times 10^{-3}$ & $\sqrt{}$ \\
{\hfill g} & 8.5 & $0.93/A^2$ & --30 & --4.7 & $-1 \times 10^{-8}$ &
$2.5 \times 10^{-3}$ & $\sqrt{}$ \\  
\hline
SUGRA {\hfill h} & $0.5$ & --11 & 1.08 & 4.4 & 
$-9 \times 10^{-7}$ &
$-0.3\times 10^{-4}$ & $\times$ \\
{\hfill i} & $0.5$ & --11 & --0.85 & 4.5 & 
$-9 \times 10^{-7}$ &
$-0.9 \times 10^{-4}$ &  $\times$ \\
{\hfill j} & 20 & --2 & 25 & 10.7 & 
$-3 \times 10^{-6}$ &
$4.4 \times 10^{-4}$ & $\sqrt{}$ \\
{\hfill k} & 2.2 & --2 & --1.7 & --0.8 & 
$-2 \times 10^{-8}$ &
$- 9 \times 10^{-5}$ &  $\sqrt{}$ \\
\hline
\end{tabular}
\caption{\label{tabela} Approximate values for $\zeta$,
  $\dot{\alpha}/\alpha(z=0)$ and $\Delta \alpha/\alpha$ for BBN $(z=10^{10})$
and Oklo phenomenon $(z=0.14)$ epochs for several quintessence models. 
2EXP: $V = V_0 \left(e^{A \kappa \phi} + e^{B \kappa \phi} \right)$,
Ref.~\cite{Barreiro:1999zs};
A-S:  $V = \kappa^{-4} e^{-A \kappa \phi}\left[ (\kappa \phi - C)^2 +
B)\right]$, Ref.~\cite{Albrecht:1999rm}; 
SUGRA: $V = V_0 \exp\left(A (\kappa \phi)^2 \right) (\kappa \phi)^B$,
  Ref.~\cite{Ng:2001hs}.  In the Sugra model (h) the initial
  condition of the field is $(\kappa \phi_{\rm in})^2 \ll B/2A$ and in 
  model (i) $(\kappa \phi_{\rm in})^2 \gg B/2A$.
The models which have late time oscillations of the
  field have a tick in the last column. We have assumed $\Omega_{\phi}
  = 0.7$ at present.}
\end{center}
\end{table*}

The SUGRA model (h) in the table is the original one of 
Ref.~\cite{Brax:1999gp},
in which the field rolls along the power law slope of the potential
whereas in the remaining cases the field starts rolling from the
exponential side. 
This table suggests that various models can be 
consistent with the present day constraints on 
$\dot\alpha/\alpha$ and with the BBN bounds, however only  a few
satisfy the Oklo bound and only at the cost of severe fine
tuning. The latter models are labeled with (c), (f), (g) and (k) 
in Table~\ref{tabela}. In these models, the field performs a low 
amplitude late time
oscillation hence satisfying the Oklo bound.
Models (f) and (g) were recently analysed in
Ref.~\cite{AG}. The possibility for the kind of behaviour of model (g) 
was pointed out before in Ref.~\cite{Chiba}. 
It is clear from Fig.~\ref{figure10} and Table \ref{tabela} that in order 
to have a realistic way of  constraining quintessence models we need to 
employ the wealth of data available on the variations of $\alpha$ over time 
scales much larger than those observed through the QSO observations.  

In the next section we propose a mechanism that enable us to reconcile
a large detection of a variation on $\alpha$ at large redshift 
with the tight Oklo constraint without need of fine tuning in the 
parameters of the potential.

\section{Non-minimal coupling of quintessence to the electromagnetic field}

In this section, we propose a radical alternative to the allowed
couplings between 
the quintessence and electromagnetic field, which will allow us to 
reconcile all the current 
constraints on the variation of $\alpha$ with the data. We start by asking 
what motivates a linear coupling between quintessence and the electromagnetic 
field? Obviously, there are two simple arguments for this choice:
its apparent simplicity and the fact that almost any interaction can be 
reduced to this form if $B_F(\phi)$ can be expanded as a series in  
$\phi(z)-\phi_0$ provided that the latter is small. 
This linearization can break down for larger $\phi(z)$, 
and thus the comparison with $\Delta \alpha /\alpha$ at very high 
$z$ (i.e. BBN limits) can only tell us whether the linear dependence 
needs to be modified. Of course the linear choice is not the only 
possibility, and quadratic couplings between matter and the scalar 
field \cite{DP,OP} and const+$\exp{\{-\kappa\phi\}}$ couplings \cite{Damour:2002nv} 
being also actively discussed in the past. 
All these approaches share the same weakness: they 
do not calculate these dependences from first principles 
but rather simply postulate them. In many ways it is analogous to simply 
choosing a potential for a tracker field by hand.
Clearly, when an {\em ad hoc} choice of  $B_F(\phi)$ 
is allowed, the profile of $\alpha(t)$ cannot be calculated and 
any $\alpha(t)$ is possible. In particular, one can ``engineer'' 
such a function $B_F(\phi)$ such that both the Oklo constraint and 
Webb {\em et al.} data points are satisfied. Moreover, one can 
choose $ d B_F(\phi)/d\phi(z=0) \ll \kappa $, 
suppressing the fifth force mediated by $\phi$ \cite{DP,Damour:2002nv}. 

We would like to point out another rather generic scenario where the 
effective fine structure constant is modified in atomic physics but
has almost no impact on nuclear and particle physics. This would 
allow us to reconcile the Oklo constraints with the results of Webb 
{\em et al.} 
and suppress the non-universality of the gravitational force 
mediated by $\phi$.
To do this we propose a momentum dependent modification of the
interaction (\ref{tx}). 
Quite generically, the vertex between two photons and a $\phi$ field can be  
written in the momentum representation as
\begin{eqnarray}
\left[ - \left(p_\mu + \fr{1}{2}q_\mu\right)
\left(p_\nu-\fr{1}{2}q_\nu\right) \right. \hspace{1.2cm} \nonumber \\
+ \left. g_{\mu\nu} \left(p^2-\fr{q^2}{4}\right) \right]
f[p^2,q^2,(pq),\phi] \,,
\label{vertex}
\end{eqnarray}
where $p\pm q/2$ are photon momenta and $q$ is the momentum of the 
scalar field. Eq.~(\ref{vertex}) has introduced a form factor $f$ that 
depends on 
the momenta 
and the value of the scalar field.
The minimal choice of coupling, $f[p^2,q^2,(pq), \phi]=
\kappa\zeta$, was discussed in previous sections. 
For all practical purposes, one can neglect the momentum of the scalar 
field so that 
\be
f[p^2,q^2,(pq)] = f[p^2] \,.
\ee
Let us now imagine that this coupling has a form factor which tends to zero 
as $p^2$ becomes larger than a certain critical momentum scale $\mu$. 
For definiteness, we assume a power-like suppression,
\be
f[p^2] = \kappa\tilde \zeta \fr{\mu^2}{\mu^2 - p^2} \,,
\label{f}
\ee
where $\tilde \zeta$ is the momentum independent 
fundamental constant. 
This choice of $f$ leads to the anomalous dependence of 
the effective coupling constant on the photon virtuality such that
\begin{eqnarray}
\alpha(p^2) = \alpha_c (1+ \kappa \zeta_{\rm eff} \phi) \hspace{0.35cm} 
{\rm with} \nonumber\\
\zeta_{\rm eff}(p^2) = \tilde \zeta ~\frac{\mu^2}{\mu^2-p^2} = 
\left\{
\begin{array}{ll}
\label{low}
\tilde \zeta \hspace{0.55cm} {\rm for} \,\, |p^2| \ll \mu^2  \\
0 \hspace{0.55cm} {\rm for} \,\, |p^2| \gg \mu^2
\label{p^2}
\end{array}
\right. \,.
\end{eqnarray}
In other words, the high virtuality photons always couple to matter
with a constant 
time-independent strength $\alpha_c$, while the couplings of low 
virtuality photons are modified by 
the presence of $\phi$ which evolves with cosmological time. 
Of course, $ \alpha_c $
will have a usual high-energy log-type running with the momentum.
In the language of operators the choice (\ref{f}) would correspond to a 
new non-local operator that can be written as 
\be
{\cal L}_{\phi F} = -\fr{1}{4}F_{\mu\nu}F^{\mu\nu}-{1\over 4}\tilde \zeta 
\kappa \phi
F_{\mu\nu}\fr {\mu^2}{\mu^2 +\partial^2}F^{\mu\nu} \,.
\label{newoper}
\ee
The zeroth order term in the expansion over $\partial^2/\mu^2$ corresponds to 
a usual linearized form of (\ref{bc}) that we were using in the
previous sections.
We also note that Eq.~(\ref{newoper}) does not allow  
for arbitrary additive redifinition of $\phi$ at $z=0$. 
As before, we assume that the cosmological evolution of $\phi$ is 
driven by its self potential. 

In the absence of the form factor, the effective coupling between 
nucleons and scalar field, $\zeta_N\kappa\phi \bar N N$ (see the notations of 
Ref. \cite{OP}), induced via 
the coupling to photons is given by 
$\zeta_{N} \sim 10^{-3} \zeta < 10^{-6}$ \cite{GL},
leading to the bound (\ref{zetabound}). 
It is clear that in order to obtain a further suppression of $\zeta_N$
coupling by the form factor (\ref{f}) one should choose the scale $\mu$ to be
lower than the characteristic virtualities of photons involved 
in the hadronic and nuclear matrix elements. The latter are of the 
order of the inverse characteristic size of a system, 
which in the case of $\zeta_N$ coupling is the size of the nucleon. 
For example, the coupling of $\phi$  to nucleons, $\zeta_N$, 
will receive a typical suppression of $\mu^2/
(200~ {\rm MeV})^2$ for $\mu\la 200~ {\rm MeV}$:
\be
\zeta_{N} \sim 10^{-3} \tilde \zeta \fr{\mu^2}{(200~ {\rm MeV})^2},
\label{suppression}
\ee
where $200~ {\rm MeV}$ is roughly the inverse size of a nucleon. 
This suppression can be even stronger for a steeper form factor. 
This leads to a considerable weakening of all constraints coming from 
the non-universality due to the $\phi$-exchange. 
For example, choosing $\mu$ 
around $10$ MeV, we get an additional $10^{-3}-10^{-2}$ suppression
so that the effective coupling to nucleons is now much smaller
$\zeta_{N} \sim (10^{-6}-10^{-5}) \tilde \zeta$. This results in a much weaker, 
$\tilde \zeta \la 1$, bound than eq. (\ref{zetabound}). 

Likewise, the bounds coming from Oklo phenomenon will also be weakened. 
This is because in the presence of the form factor (\ref{p^2})
the dependence of nuclear levels on $\phi$ entering in the standard 
calculation \cite{DP,Fujii,OPQ} is suppressed and consequently 
these levels are less affected by the change in the $\phi$ field. 
Indeed, the expected weakening 
of the Oklo bound is at least a factor of $\mu^2/(50~ {\rm MeV})^2\sim 0.04$ 
with the same choice of $\mu$. Here $50 $ MeV is a typical inverse radius 
of a heavy nucleus. This is just enough to make the results of 
Oklo analysis \cite{DP,Fujii,OPQ} compatible with Webb {\em et al.} 
even with $\dot \alpha \simeq$ const and {\em without any fine tuning}. 

On the other hand, the atomic physics phenomena 
for which the characteristic photon virtualities 
are much smaller than $\mu \sim 10$ MeV will be sensitive to
``dressed'' $\alpha$ that evolves as a function of $\phi$. 
In particular, this refers to the relativistic corrections 
to the atomic energy levels, used for the QSO analysis of Webb et al. 
Thus, the existence of the form factor 
allows to ``isolate'' the phenomenon of changing alpha to atomic physics and 
remove it from nuclear and particle physics.

The anomalous dependence of the fine structure constant on the photon momentum 
(\ref{low}) can be probed 
with high precision QED measurements, as different experiments 
probe different photon virtualities. Some of these measurements are 
performed with atomic and condensed matter systems, where the 
characteristic momentum transfers are much smaller than $\mu$. 
Take, for example, the determination of $\alpha$ from the 
quantum Hall effect where the characteristic momenta of photons are very small.
This experiment will be sensitive to the ``dressed'' 
value of $\alpha(p^2\simeq 0)$ with obvious condition 
$\phi=\phi(z=0)$. 
On the other hand, the precision measurements sensitive to  
QED radiative corrections such as electron and muon anomalous magnetic moments 
can also probe $\alpha$ at virtualites comparable to $\mu$ where 
the value of $\alpha$ 
is given by (\ref{p^2}). The expression for the magnetic dipole moment 
of the electron and muon including the first loop correction will be given by
\be
d_i \simeq \fr{e(p^2\simeq 0)}{m_i}\left[1+ 
\fr{\alpha(p^2 \simeq m_i^2)}{2\pi}\right] \,,
\label{d_i}
\ee
where $i = e,\mu$.
The electron $g-2$ experiment \cite{g-2e}
has measured the anomalous magnetic moment of the 
electron to an accuracy that allows for the extraction of $1/\alpha$ 
to nine digits. 
Now we have to take into account that the one-loop correction has some 
photons with virtualities $p^2\simeq \mu^2 \sim (10 ~{\rm MeV})^2$.
This will introduce a correction to $g-2$  
of the electron due to the anomalous 
behaviour of $\alpha$ at large momenta at the level of 
\begin{eqnarray}
\Delta (g-2)_e &\sim & \tilde \zeta \kappa\phi \left(\frac{\alpha}{2\pi}\right)
\left(\frac{m_e^2}{\mu^2}\right) \nonumber \\
&\sim&  10^{-10}-10^{-11} \,,
\label{ge}
\end{eqnarray}
for $\mu\sim 10$ MeV and typical $\tilde \zeta\kappa \phi \sim 10^{-5}$ 
suggested by Webb {\em et al.} results. This correction is 
taken relative to the value of $g-2$ calculated with ``fully dressed'' 
value of 
$\alpha$ (\ref{low}).  Since $(g-2)_e$ is used 
for the extraction of $\alpha$, one should absorb correction (\ref{ge})
into the measured value of $\alpha$ and compare it with the second 
best determination of the fine structure constant from the quantum 
Hall effect. The correction (\ref{ge}) corresponds to the shift 
of $1/\alpha$ in 
the eighth digit 
which appears to be marginally consistent with the determination of $\alpha$ 
from the solid state physics \cite{Mohr:2000ie}. 

The largest effect is expected in the $g-2$ of the muon, where the radiative 
corrections probe ``undressed'' $\alpha$, or $\alpha_c$, (\ref{p^2})
at virtualities comparable to the muon mass,
which is much larger than a chosen value of $\mu\simeq 10 $ MeV. 
Therefore, the expected correction to the muon anomalous 
magnetic moment (relative to the value one would expect with 
``dressed'' $\alpha$ given by (\ref{low})) will be on the order
\be
\Delta (g-2)_\mu \sim \tilde \zeta \kappa\phi \left(\frac{\alpha}{2\pi}\right),
\label{gmu}
\ee
which is {\em larger} than both the experimental and theoretical 
accuracy of $(g-2)_\mu$, unless
$\tilde \zeta \kappa\phi < 10^{-6}$, where $\phi$ is the value of the 
scalar field {\em now}. 
This might still be marginally consistent with Ref.~\cite{Webb01} if 
$\tilde \zeta \kappa\phi(z\simeq 0.5) \sim 10^{-5}$. 

Therefore, the anomalous running of $\alpha$ with the momentum 
(\ref{low}), (\ref{p^2}) comes tantalizingly close to the accuracy 
of modern tests of QED. At the same time, it allows us 
to disconnect the phenomenon of ``changing $\alpha$'' from 
nuclear and hadronic physics. 
This way the results of Webb {\em et al.} coming from the 
atomic physics, are not in 
contradiction with Oklo phenomenon and are of no immediate consequence for the 
fifth-force experiments. 

\section{Conclusions}
We have studied a variety of cosmological scalar field models coupled 
to the electromagnetic Lagrangian. The evolution of the scalar field 
over cosmological times induces the effective change of the fine 
structure constant.
We have considered two generic possibilities:

(i) Tracker solution.  We have shown that for a wide range of models where the
scalar field is tracking, there exists a region in parameter space which allows
concordance with the QSO data on non-zero $\Delta\alpha/\alpha$ 
\cite{Webb01} with other constraints on $\Delta \alpha/\alpha $ 
coming from the Big Bang nucleosynthesis as well as the precision 
measurements of the universality of the gravitational force.
The predictions for the present day $\dot\alpha/\alpha$ are typically within 
one order of magnitude from current experiments and within the reach of
near future laboratory experiments. 
Typical constraints on the parameter space are 
shown in  Figs.~\ref{figure3} and \ref{figure51}. 
There is an upper bound on the mass scale $M$ or $V_0$ from the 
requirement that the tracking solution persists 
until $z=0$ and  $\rho_{\phi} \ll \rho_B^{}$ today. 
The consequence of this is the lower bound on $\zeta$, 
the coupling between the scalar field and the electromagnetic 
$F_{\mu\nu}F^{\mu\nu}$. This typically leads to $\zeta > 10^{-5}$ (see
Figs.~\ref{figure3}, \ref{figure51} and from the dashed line in
Fig.~\ref{omegadm}) in the region of parameters studied.  
The universality of the gravitational force puts an upper 
bound on the coupling, $\zeta < 10^{-3}$.

(ii) Dark energy. We have found that the simple inverse power law
model for quintessence also offers a {\it small} region of parameter space in
agreement with the QSO observations. Here, for a chosen $n$ 
the mass scale $M$ is fixed,
therefore, the only free parameter remaining is the 
coupling $\zeta$. 
To fit the QSO data for a chosen $n$ we obtain one single
$\zeta$ (or a small window due to the uncertainties related with the data).
We have also considered the model of the scalar field coupled to
the dark matter. We have found that only large values (if $\lambda$ is
negative) of the slope of the scalar
potential allow this model to pass the equivalence principle
constraints. 
This model can generate oscillatory behaviour for $\alpha(z)$ which 
may lead to the 
enhancement of the present day signal of $\dot \alpha/\alpha$. 
We looked at other
models of quintessence suggested in the literature. Their relevant
properties for this work are summarised in Table \ref{tabela}. 
We expect that complimentary information on the present equation of
state of the universe, on the coupling $\zeta$ and on the ratio 
$\dot{\alpha}/\alpha$ can help us to
back up or rule out models of quintessence and scenarios in which a
dynamical subdominant scalar field is evolving in a tracking regime.

A notable feature of this detailed analysis is that 
apart from a small region of parameter
space in some quintessence models with late time oscillations 
(e.g., models (c), (f), (g) and (k)
of Table~\ref{tabela}) 
we have not come accross a model that satisfies both 
the Oklo bound $|\Delta \alpha/\alpha| \la 10^{-7}$ at $z \sim 0.14$ and 
$\Delta \alpha/\alpha \sim -10^{-5}$ at 
$z \sim 1$ reported by Webb {\it et al.}. 
In an attempt to reconcile this, we have suggested that the existence of  
form factor in the 
coupling of $\zeta$ at around 10 MeV with respect to the photon 
momentum may lead to the 
suppression of the effective coupling between $\phi$ and nucleons and nuclei. 
This is sufficient to isolate the phenomenon of changing couplings 
to the realm of atomic physics, while relaxing the bounds coming from the 
Oklo phenomenon and the fifth-force experiments. This amounts 
to postulating an anomalous running of $\alpha$ with momentum 
and thus gives an additional possibility of checking this idea with 
high-precision QED measurments, sensitive to various ranges of the 
photon virtualities.

\begin{acknowledgements}
Two of the authors, NJN and MP acknowledge PPARC for financial support. 
The research of MP is supported in part by NSERC of Canada. 
 \end{acknowledgements}

\end{document}